\documentclass[aps,prd,reprint,groupedaddress]{revtex4-1}
\pdfoutput=1 

\usepackage{blindtext}
\usepackage{hyperref}
\usepackage{mathtools, cases}
\usepackage{amsfonts}
\usepackage[T1]{fontenc} 
\usepackage{float}
\usepackage[export]{adjustbox}

\newcommand{\bx}[1]{\mathbf{x}_{#1}}
\newcommand{\bw}[1]{\mathbf{w}_{#1}}
\newcommand{\cmbjl}{\textsc{CMBLensing.jl}}

\newcommand{\RUG}{Kapteyn Astronomical Institute, University of Groningen, 9700~AV Groningen, The Netherlands}
\newcommand{\VSI}{Van Swinderen Institute for Particle Physics and Gravity,\\ University of Groningen,
Nijenborgh 3, 9747 AG Groningen, The Netherlands}
\newcommand{\CCA}{Center for Computational Astrophysics, 162 5th Avenue, New York, NY, 10010, USA}
\newcommand{\kavli}{Kavli Institute for Cosmology, Madingley Road, Cambridge, CB3 0HA, UK}
\newcommand{\damtp}{DAMTP, Centre for Mathematical Sciences, Wilberforce Road, Cambridge CB3 0WA, UK
}
\newcommand{\paris}{Sorbonne Universit\'{e}, CNRS, UMR 7095, Institut d’Astrophysique de Paris, 98 bis bd Arago, 75014 Paris, France}
\newcommand{\princeton}{Department of Astrophysical Sciences, Princeton University, 4 Ivy Lane, Princeton, NJ 08544 USA}

\begin{document}
\title{Denoising Diffusion Delensing Delight: \\Reconstructing the Non-Gaussian CMB Lensing Potential with Diffusion Models}

\author{Thomas Fl\"oss\textsuperscript{1,2}}
\email{tsfloss@gmail.com}
\author{William R. Coulton\textsuperscript{3,4}}
\author{Adriaan J. Duivenvoorden\textsuperscript{5}}
\author{Francisco Villaescusa-Navarro\textsuperscript{5,6}}
\author{Benjamin D. Wandelt\textsuperscript{7,5}}

\affiliation{\textsuperscript{1}\VSI}
\affiliation{\textsuperscript{2}\RUG}
\affiliation{\textsuperscript{3}\kavli}
\affiliation{\textsuperscript{4}\damtp}
\affiliation{\textsuperscript{5}\CCA}
\affiliation{\textsuperscript{6}\princeton}
\affiliation{\textsuperscript{7}\paris}

\begin{abstract}
Optimal extraction of cosmological information from observations of the Cosmic Microwave Background critically relies on our ability to accurately undo the distortions caused by weak gravitational lensing. In this work, we demonstrate the use of denoising diffusion models in performing Bayesian lensing reconstruction. We show that score-based generative models can produce accurate, uncorrelated samples from the CMB lensing convergence map posterior, given noisy CMB observations. To validate our approach, we compare the samples of our model to those obtained using established Hamiltonian Monte Carlo methods, which assume a Gaussian lensing potential. We then go beyond this assumption of Gaussianity, and train and validate our model on non-Gaussian lensing data, obtained by ray-tracing N-body simulations. We demonstrate that in this case, samples from our model have accurate non-Gaussian statistics beyond the power spectrum. The method provides an avenue towards more efficient and accurate lensing reconstruction, that does not rely on an approximate analytic description of the posterior probability. The reconstructed lensing maps can be used as an unbiased tracer of the matter distribution, and to improve delensing of the CMB, resulting in more precise cosmological parameter inference.

\end{abstract}

\maketitle

\section{Introduction}
The deflection of photons of the Cosmic Microwave Background (CMB) by the large-scale structure between today and the last-scattering surface, changes the CMB statistics, leaving a characteristic imprint \cite{Cole:1989,Linder:1990,Seljak:1995ve,Metcalf:1997ih,Lewis:2006fu}. This effect, known as weak gravitational lensing, affects our ability to infer properties of the early universe such as primordial gravitational waves through B-mode polarization \cite{Seljak:2003pn}, or the presence of primordial non-Gaussianity \cite{Babich:2004yc,Coulton:2019odk, Coulton:2022wln}, both of which would constitute strong evidence for the inflationary paradigm and are thus important targets for experimental endeavors \cite{Achucarro:2022qrl}. Furthermore, the gravitational potential that lenses the CMB is an unbiased tracer (albeit projected) of the total matter distribution of the universe and thus contains important cosmological information by itself.
Accurately reconstructing the CMB lensing potential is therefore of considerable interest \cite{Green:2016cjr,Hotinli:2021umk,Trendafilova:2023xtq,Ange:2023ygk}, for example, to constrain the mass of neutrinos \cite{TopicalConvenersKNAbazajianJECarlstromATLee:2013bxd}, the late-time growth of structure (i.e $\sigma_8$) and local primordial non-Gaussianity, especially in cross-correlation with other tracers, through cosmic-variance cancellation \cite{Seljak:2008xr,Schmittfull:2017ffw}.\\

Several methods have been developed to reconstruct the lensing potential from observed CMB data, starting with the quadratic estimator by Seljak \& Zaldarriaga \cite{Zaldarriaga:1998te}, and Hu \& Okamoto \cite{Hu:2001kj,Okamoto:2003zw}. These estimators have been used in virtually all CMB lensing analyses to date \cite{Das:2011ak,vanEngelen:2012va,POLARBEAR:2013oat,SPT:2014puc,Planck:2013mth,Planck:2015mym}, for which they proved to be sufficiently optimal. The high resolution and low noise levels of current (South Pole Telescope 3G \cite{SPT-3G:2014dbx,SPT-3G:2024qkd}) and next generation (Simons Observatory \cite{SimonsObservatory:2018koc} and CMB-S4 \cite{CMB-S4:2016ple}) CMB surveys, allow sub-percent level lensing reconstruction and an unprecedented sensitivity to primordial B modes. At this point, however, the quadratic estimator ceases to be nearly optimal, and more sophisticated techniques are required to achieve optimal lensing reconstruction, such as iterative maximum a posteriori (MAP) estimators \cite{Hirata:2002jy,Hirata:2003ka,Carron:2017mqf,Millea:2017fyd} and gradient inversion techniques \cite{Hadzhiyska:2019cle}. Besides these analytical approaches, the use of machine learning methods in CMB lensing problems has been an active topic of research in recent years as well \cite{Caldeira:2018ojb,Guzman:2021nfk,Li:2022kbd, Yan:2023irb}. Although the quality of the reconstructed lensing potential with iterative or machine-learned estimators may be statistically optimal, both methods yield point estimates that require expensive simulations to propagate uncertainty in the estimates when they are used in any subsequent analyses (e.g. to infer primordial non-Gaussianity from delensed CMB data). Additionally, determining non-Gaussian statistics of the lensing potential using these point estimates can be challenging, due to having to model complicated noise biases \cite{Kalaja:2022xhi}, although this may be somewhat alleviated with alternative estimators \cite{Namikawa_2013}. \\

More recently the problem of CMB delensing and lensing reconstruction has received a fully Bayesian treatment by Millea, Anderes \& Wandelt \cite{Millea:2017fyd,Millea:2020cpw}, which we refer to as MAW from hereon out (see also Ref.~\cite{Anderes:2014foa} for earlier work using only temperature modes). Their approach to this high-dimensional statistical problem relies on having access to the gradient of the CMB lensing posterior, allowing for the use of a Hamiltonian-Monte-Carlo-within-Gibbs sampler that simultaneously samples cosmological parameters (the amplitude $A_\phi$ of the lensing potential power spectrum, and the tensor-to-scalar ratio $r$), the delensed CMB data and the lensing potential. The ability to sample from the posterior enables the accurate determination of uncertainties when working with the delensed data or the inferred lensing potential. This powerful method has since been used to improve the analysis of data from SPT-3G \cite{Millea:2020iuw}.\\

One of the crucial assumptions for MAW being able to write down and evaluate the posterior is that the lensing potential is Gaussian. However, in reality, this assumption does not hold since the large-scale structure of the universe is non-Gaussian. More accurate reconstruction of the lensing potential will thus require going beyond this Gaussian \emph{prior} assumption, but this is not straightforward with existing methods. At the same time, the Hamiltonian Monte Carlo (HMC) method typically requires long sampling chains to ensure convergence of the lensing maps, which scales inefficiently with the size of the data. Although this issue can be partially ameliorated with suitable approximations \cite{millea2022improved,Millea:2021had}, it is worth investigating alternative \emph{simulation-based inference} solutions \cite{Cranmer:2019eaq,Alsing}. \\

Here, we explore the use of probabilistic machine learning models in reconstructing the lensing potential, alleviating some of the limitations of the HMC approach.  Specifically, we consider the use of score-based generative models (SGM) \cite{Song2020}, a variant of what is commonly referred to as \emph{diffusion models}. Using techniques inspired by the stochastic nature of diffusion, these models can draw uncorrelated samples from complicated high-dimensional probability distributions, without requiring an explicit expression for this distribution, making them particularly suitable for problems involving images \cite{ho2020denoising}. Simply put, such models learn to map samples from a simple distribution (i.e. a Gaussian) into samples of a more complicated distribution, iteratively turning pure noise into meaningful data. Diffusion models have been successfully applied to various problems in the context of astrophysics and cosmology (e.g. \cite{Remy:2022ixn,Adam:2022zji,Legin:2023jxc,Karchev:2022ycy,Ono:2024jhn,Schanz:2023uzg,Rouhiainen:2023ewv,Heurtel-Depeiges:2023huy}). Here we will demonstrate the use of such models in reconstructing the CMB lensing convergence map given noisy CMB observations, enabling sampling of the Bayesian lensing posterior. While MAW solves the problem with explicitly specified likelihood and priors, our method implements an implicit inference approach where likelihood and prior are represented in terms of pairs of lensing and data maps. In our case these are generated through simulations, making our work an example of simulation-based inference. Because of this, we are able to reconstruct non-Gaussian lensing potentials with accurate non-Gaussian statistics for the first time, requiring only a change in training data. Our model, trained on non-Gaussian lensing maps from N-body simulations, generates samples with the correct non-Gaussian statistics, as demonstrated by the one-point PDF and bispectrum. On signal-dominated scales, the sampled maps show excellent agreement with the true non-Gaussian lensing potential, while on noise-dominated scales, they revert to the learned non-Gaussian prior. Notably, models trained only on Gaussian lensing maps fail to capture the full non-Gaussian structure. Furthermore, our method allows for fast and uncorrelated sampling of the lensing potential posterior, circumventing the convergence issues and inefficiencies faced by traditional Markov Chain Monte Carlo methods. These advancements open up new possibilities for efficient delensing and precise constraints on large-scale structure from upcoming CMB surveys.\\

The paper is organized as follows. In section \ref{sec:lensing} we go over the necessary basics of CMB lensing. In section \ref{sec:Bayes} we introduce the Bayesian approach to CMB delensing by MAW. In section \ref{sec:SGM} we introduce some basic aspects of score-based generative diffusion models. To validate the use of these models for Bayesian lensing reconstruction tasks, we perform a detailed comparison against the method and code of MAW in section \ref{sec:benchmark}. Finally, we demonstrate that our model can be used for the accurate reconstruction of non-Gaussian lensing potentials in section \ref{sec:NG}.

\section{CMB lensing basics}
\label{sec:lensing}
The intervening matter distribution gravitationally bends the paths of CMB photons that travel through it, such that we observe deflected photons. This induces a remapping of the true CMB temperature $(T)$ and polarization (Stokes $Q$,$U$) modes to the observed, lensed quantities $(\tilde{T},\tilde{Q},\tilde{U})$ (for simplicity neglecting a small phase correction \cite{Challinor:2002cd}):
\begin{eqnarray}
\label{eq:lensed}
    \tilde{T}(\hat{\mathbf{n}}) &=& T(\hat{\mathbf{n}} + \boldsymbol{\alpha}(\hat{\mathbf{n}}))\nonumber \\
    (\tilde{Q} \pm i \tilde{U})(\hat{\mathbf{n}}) &=& (Q \pm i U)(\hat{\mathbf{n}} + \boldsymbol{\alpha}(\hat{\mathbf{n}}))
\end{eqnarray}
where $\hat{\mathbf{n}}$ is the unit vector that denotes the position in the sky, and $\boldsymbol{\alpha}$ is the deflection angle that encodes the remapping. This deflection angle is given by the gradient of the lensing potential $\phi$:
\begin{eqnarray}
    \boldsymbol{\alpha} = \nabla \phi,
\end{eqnarray}
which itself is a two-dimensional projection of the three-dimensional gravitational potential $\psi$:
\begin{eqnarray}
    \phi(\hat{\mathbf{n}}) = -2 \int d\chi \; \frac{\chi_{\mathrm{CMB}} - \chi}{\chi_{\mathrm{CMB}} \chi }\psi(\hat{\mathbf{n}}\chi,\chi),
\end{eqnarray}
where $\chi$ denotes the comoving distance, and $\chi_{\mathrm{CMB}}$ is the comoving distance to the last-scattering surface, where the CMB was emitted (at redshift $z \approx 1100$). In the flat sky approximation, these fields can also be written in Fourier space, parametrized by wavevector $\boldsymbol{\ell}$. Another useful quantity is the the CMB lensing convergence $\kappa$:
\begin{eqnarray}
   \kappa(\hat{\mathbf{n}}) = - \frac{1}{2}  \nabla^2 \phi(\hat{\mathbf{n}}),
\end{eqnarray}
which is most easily obtained in Fourier space as:
\begin{eqnarray}
\label{eq:kappaofphi_ell}
    \kappa_{\boldsymbol{\ell}} = \frac{|\boldsymbol{\ell} |^2}{2} \phi_{\boldsymbol{\ell}}.
\end{eqnarray}
Using a Taylor expansion of Eq. \eqref{eq:lensed} one can determine the lowest-order effect of lensing on the CMB fluctuations. This can then be used to construct a quadratic estimator for the CMB lensing convergence \cite{Zaldarriaga:1998te,Hu:2001kj,Okamoto:2003zw}. Additionally, iterating this estimator yields the maximum-a-posteriori estimate of the lensing convergence \cite{Hirata:2002jy,Hirata:2003ka,Carron:2017mqf}.

\section{Bayesian delensing}
\label{sec:Bayes}
In Refs. \cite{Millea:2017fyd,Millea:2020cpw}, MAW present a Bayesian approach to lensing reconstruction and delensing of the CMB fluctuations. The approach takes as a starting point the posterior probability distribution of the true CMB $f$ (i.e. $T,Q,U$), the lensing potential $\phi$ and cosmological parameters $\theta$ (the amplitude of the lensing power spectrum $A_\phi$ and the tensor-to-scalar ratio $r$), given observed data $d$ (i.e. $\tilde{T},\tilde{Q},\tilde{U}$): $\mathcal{P}(f,\phi,\theta | d)$. This data constitutes lensed, noised, and masked CMB data:
\begin{eqnarray}
    d = \mathbb{A}\,\mathbb{L}(\phi) f + n
\end{eqnarray}
where $\mathbb{L}(\phi)$ denotes the lensing operation, $n$ is the noise, and $\mathbb{A}$ is a linear transformation that includes instrumental effects such as the beam and sky mask. Using Bayes' theorem, this posterior can be expressed in terms of the likelihood of the data, and a prior (following the concise notation of MAW):
\begin{eqnarray}
    \mathcal{P}(f,\phi,\theta | d) \propto \mathcal{P}(d | f,\phi,\theta) \mathcal{P}(f,\phi,\theta).
\end{eqnarray}
Under the assumption that the noise is a Gaussian random field with covariance $\mathbb{C}_n$, the likelihood is proportional to:
\begin{eqnarray}
    \mathcal{P}(d | f,\phi,\theta) \propto \exp{\left(-\frac{(d - \mathbb{A}\,\mathbb{L}(\phi) f)^2}{2 \mathbb{C}_n}  \right)},
\end{eqnarray}
and the priors on the $f$ and $\phi$ fields are taken to be Gaussian:
\begin{eqnarray}
    \mathcal{P}(f,\phi,\theta) \propto \exp{\left(-\frac{f^2}{2 \mathbb{C}_f}  \right)}\exp{\left(-\frac{\phi^2}{2 \mathbb{C}_\phi}  \right)} \mathcal{P}(\theta).
\end{eqnarray}
Since we are after the pixels of the fields $f$ and $\phi$, this constitutes a high-dimensional posterior distribution that is intractable with conventional Metropolis-Hastings Monte Carlo methods. MAW overcome this challenge by numerically implementing the posterior in a fully differentiable fashion (see \cmbjl{} \footnote{\url{https://github.com/marius311/CMBLensing.jl}}), providing access to the derivatives of the posterior with respect to all parameters (i.e. field pixels and $\theta$). This enables both MAP estimation through gradient-based optimization \cite{Millea:2017fyd} and Hamiltonian Monte Carlo sampling \cite{Millea:2020cpw} of the posterior. Crucially, even if one is only after the lensing potential, they find that it is easier to sample the full posterior $\mathcal{P}(f,\phi,\theta|d)$, rather than the marginalized posterior $\mathcal{P}(\phi,\theta|d)$.\\

Although the Gaussian prior on the lensing potential does not imply that the posterior samples are necessarily Gaussian, it is not the maximally informative prior for data that have been lensed with a non-Gaussian potential, and we expect samples in this case to have inaccurate non-Gaussian statistics. Including such a non-Gaussian prior in the approach of MAW would require forward modeling and evaluation of the posterior, which is expensive and hard to implement. Additionally, HMC methods can be inefficient, due to subsequent samples being correlated. In practice, the sample chains of the large-scale modes of the lensing potential have long auto-correlation lengths, resulting in a low yield of independent samples of these scales, especially when a mask is included \cite{Millea:2020cpw,Carron:2017mqf}.\\

In the rest of this work, we will demonstrate the use of generative machine learning models in reconstructing the CMB lensing convergence to overcome both of these limitations of the HMC approach, albeit at the expense of losing provable long-run convergence to the true posterior.

\section{Score-based generative models}
\label{sec:SGM} 
For our probabilistic machine learning approach, we adopt score-based generative models \cite{Song2020}. These models can produce samples of a learned data distribution $p(\bx{})$ through a reverse-diffusion process. Starting from data $\bx{}$, we incrementally perturb the data using Gaussian noise with an increasing variance schedule $\sigma^2(t)$ over many timesteps $t \in [0,1]$, until it is pure noise. This \emph{diffusion} process is described by a stochastic differential equation (SDE) \cite{Song2020}:
\begin{equation}
    \label{eqn:fwdsde}
    d\bx{} = \mathbf{f}(\bx{},t)dt + g(t) d\bw{},
\end{equation}
where $\mathbf{f}$ is called the drift coefficient, $g$ is the diffusion coefficient, and $d\bw{}$ denotes a Wiener (noising) process. Remarkably, the diffusion process has a reverse SDE that turns pure Gaussian noise into a sample of $p(\bx{})$ by reversing the noising process at each timestep; the \emph{denoising} process \cite{anderson1982reverse}:
\begin{equation}
    \label{eqn:bwdsde}
    d\bx{} = \left[\mathbf{f}(\bx{},t) -g(t)^2\nabla_{\bx{}}\log{p_t(\bx{}})\right]dt + g(t) d\bar{\bw{}}.
\end{equation}
This requires access to the \emph{score} $\nabla_{\bx{}}\log{p_t(\bx{})}$, where $p_t(\bx{})$ is the probability distribution of $\bx{}$ at time $t$. Furthermore, if we are after the conditional distribution $p(\mathbf{x}|\mathbf{y})$ (i.e. the posterior of parameters $\mathbf{x}$ given data $\mathbf{y}$), we instead need the conditional score $\nabla_{\bx{}}\log{p_t(\mathbf{x}|\mathbf{y})}$. In our case $\mathbf{x}$ represents the lensing convergence map, while $\mathbf{y}$ represents observed lensed CMB data (e.g. $Q$ and $U$ maps). We use a neural network, more specifically a U-Net, to approximate this conditional score by training it to predict random noise added to an image $\mathbf{x}$ at a random timestep, given conditioning data $\mathbf{y}(\mathbf{x})$ \cite{batzolis2021conditional}.\\

Typically in diffusion models, the noising process $d\mathbf{w}$ uses white noise
\begin{eqnarray}
    \mathbf{z}_{\rm{w}} \sim \mathcal{N}(\mathbf{0},\mathbf{I}),
\end{eqnarray}
but we find that both training and sampling are faster and more accurate when using \emph{reddened} Gaussian noise with a power spectrum similar to the target $\kappa$ maps:
\begin{eqnarray}
    \mathbf{z}_{\kappa}(\mathbf{z}_{\rm{w}}) =  \mathbf{\Sigma}^{1/2}_{\kappa} \, \mathbf{z}_{\rm{w}},
\end{eqnarray}
where $\mathbf{\Sigma}_{\kappa}$ is the theory covariance matrix of the lensing convergence maps, and the multiplication is performed in Fourier space, where this covariance matrix is diagonal (i.e. the power spectrum $C^{\kappa \kappa}_{\ell}$). This type of noise allows the model to affect all scales of the data at every timestep, instead of different scales at different timesteps, resulting in higher-quality samples with fewer diffusion steps. Furthermore, we employ the Variance Exploding variant of the algorithm \cite{Song2020}:
\begin{eqnarray}
    \mathbf{f}(\bx,t) &=& 0, \hspace{1cm}
    g(t) = \sqrt{\frac{d\sigma^2(t)}{dt}}, \nonumber \\
    \sigma(t) &=& \sigma_{\rm{min}}\left( \frac{\sigma_{\rm{max}}}{\sigma_{\rm{min}}}\right)^t,
    \label{eq:schedule}
\end{eqnarray}
where $\sigma_{\rm min}$ and $\sigma_{\rm max}$ set the lowest and highest noise scales of the noising process. Given this setup, we use the noise schedule $\sigma_{\rm min} = 0.01$ and $\sigma_{\rm max} = 100$. This provides sufficient noise to properly swamp the image, as well as sufficient low-noise steps to fine-tune the sample. Finally, to not underemphasize small scales in the model's training objective, the score-matching loss to be optimized by the network is still defined at the level of the white noise:
\begin{eqnarray}
    L = \mathbb{E}_{t,\mathbf{x}_0,\mathbf{z}_{\rm w}}|| \mathbf{z}_{\rm{w}} - \mathbf{s}_\theta(\mathbf{x}_0 + \sigma(t) \mathbf{z}_{\rm{\kappa}}(\mathbf{z}_{\rm{w}}),t)||^2
\end{eqnarray}
where $\mathbf{s}_\theta(\mathbf{x}_t, t)$ is the U-Net's noise prediction at time $t$, given the noisy image $\mathbf{x}_t$.\\

After training, we sample from the learned conditional distribution by discretizing the reverse SDE with the Euler-Maruyama method:
\begin{eqnarray}
    \mathbf{x}_{t-\Delta t} = \mathbf{x}_t - g(t)^2 \left(\mathbf{\Sigma}^{1/2}_{\kappa} \; \mathbf{s}_\theta(\mathbf{x}_t, t)\right)\Delta t + g(t)\sqrt{\Delta t} \, \mathbf{z}_{\kappa},\nonumber \\
\end{eqnarray}
and applying this equation iteratively starting from pure noise at time $t=1$. We use $1000$ timesteps to draw samples, but good results can be achieved with fewer steps.\\

More details on the implementation of the model and U-Net can be found in Appendix \ref{App:UNET}.

\section{Gaussian lensing reconstruction}
\label{sec:benchmark}
To validate our method, we first perform a benchmark comparison against the HMC algorithm, using \cmbjl{}. As noted, this algorithm can be used to simultaneously sample the true CMB, the lensing potential $\phi$, and parameters $A_\phi$ (the amplitude of the lensing potential power spectrum) and $r$ (the tensor-to-scalar ratio). In this work, we will only focus on reconstructing the lensing potential, thus implicitly marginalizing over the true CMB. Ideally, we would marginalize over both the amplitude and tensor-to-scalar ratio. However, we found that the score-based generative algorithm, as well as variants thereof (e.g. Denoising Diffusion Probabilistic Models), struggle to yield samples with the correct amplitude if we vary the lensing amplitude. We comment more on this in Appendix \ref{app:amp}. For now, we limit our setup to fixed amplitude $A_\phi = 1$, but we do marginalize over $r$. \\

We mimic one of the survey configurations in MAW's work and use \cmbjl{} to generate training data, consisting of $32768$ pairs of Gaussian $\kappa$ and lensed $(Q, U)$-maps, including masks and noise with specifications given in Table \ref{tab:config} and random $r \in [10^{-6},10^{-1}]$ (quadratically spaced). The ($Q,U,\kappa$)-maps are each normalized by the mean and standard deviation over the entire dataset. Furthermore, we apply a cutoff $\ell < 3000$ to the $\kappa$ maps during training, so the network only learns to reconstruct modes below this cutoff (the $(Q,U)$ maps still have $\ell < 5000$). In our setup, modes beyond this cutoff are severely noise-dominated in the reconstruction samples and therefore not informative. Finally, since these maps have been generated with periodic boundary conditions, we apply periodic padding in our U-Net.\\

\begin{table}
\centering
\begin{tabular}{l|c}
\hline
\textbf{Parameter} & \textbf{Configuration} \\
\hline
CMB data & $Q$,$U$ \\
Map size (pixels) & $256\times 256$ \\
Pixel width & $2$ arcmin \\
Total area & $73$ deg$^2$ \\
White noise level in \( P \) & $1$ \(\mu K\)-arcmin \\
$(\ell_{\rm{knee}}$, $\alpha_{\rm{knee}})$ & $(100,3)$ \\
Beam FWHM & $2$ arcmin \\
Fourier masking  & $2 < \ell < 5000$ \\
Pixel masking  & $0.4^{\circ}$ border $+ 0.6^{\circ}$ apod \\
\hline
\end{tabular}
\caption{Configuration used in our experiments. The setup follows that of the 2PARAM setup of \cite{Millea:2020cpw}.}
\label{tab:config}
\end{table}

\begin{figure}
    \centering
    \includegraphics[scale=.355]{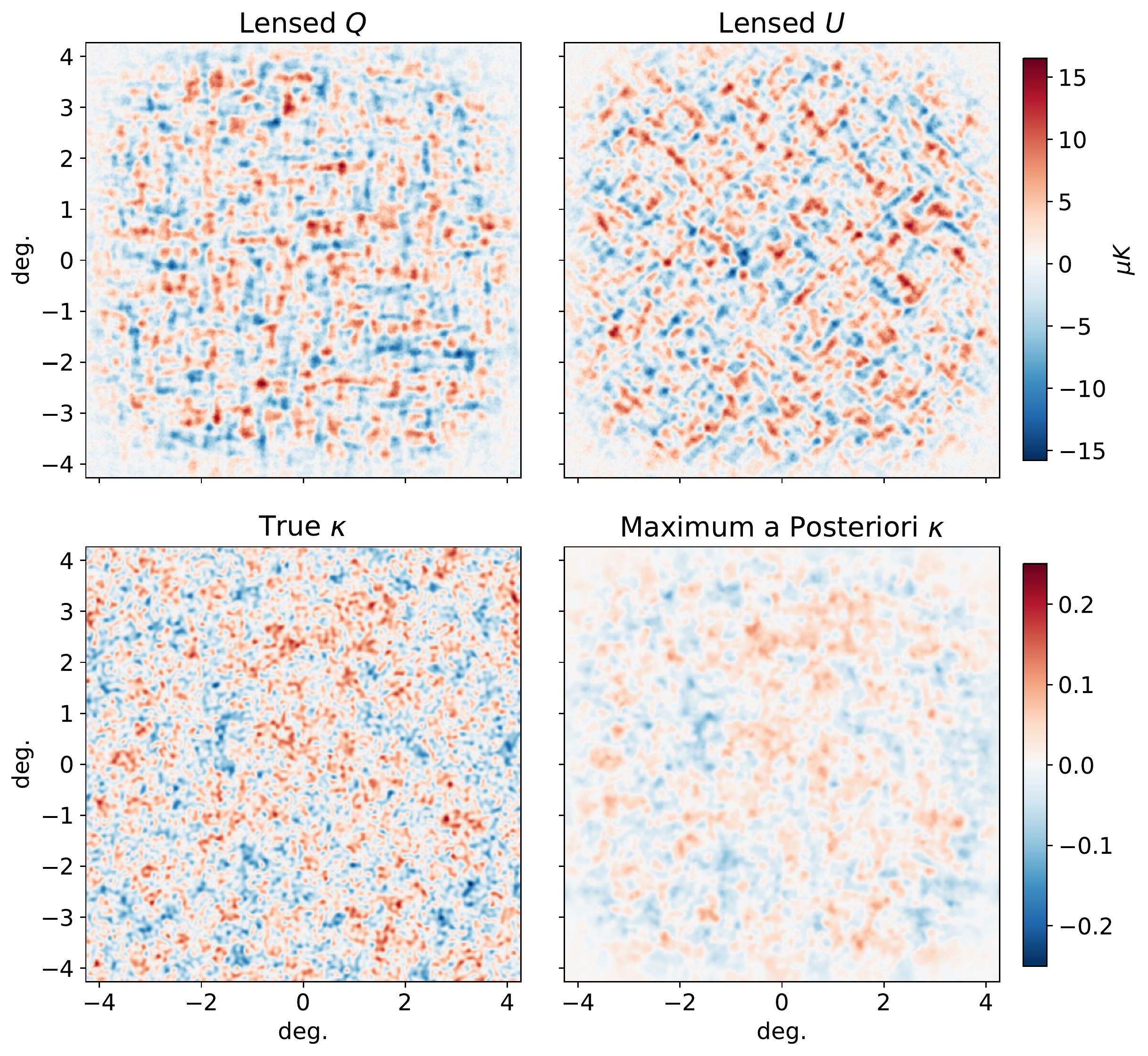}
    \caption{Validation data used in the \cmbjl{} benchmark following the configuration presented in Table \ref{tab:config}. \textbf{Top}: the lensed $(Q,U)$ observation. \textbf{Bottom left}: the true lensing convergence map $\kappa$ that lensed the $(Q,U)$ observation. \textbf{Bottom right}: the maximum a posteriori (MAP) estimate of the lensing convergence given the $(Q,U)$ observation, using the optimization algorithm of \cmbjl{}.}
    \label{fig:validation_maps_1PARAM}
\end{figure}

\begin{figure}
    \centering
    \includegraphics[scale=.355]{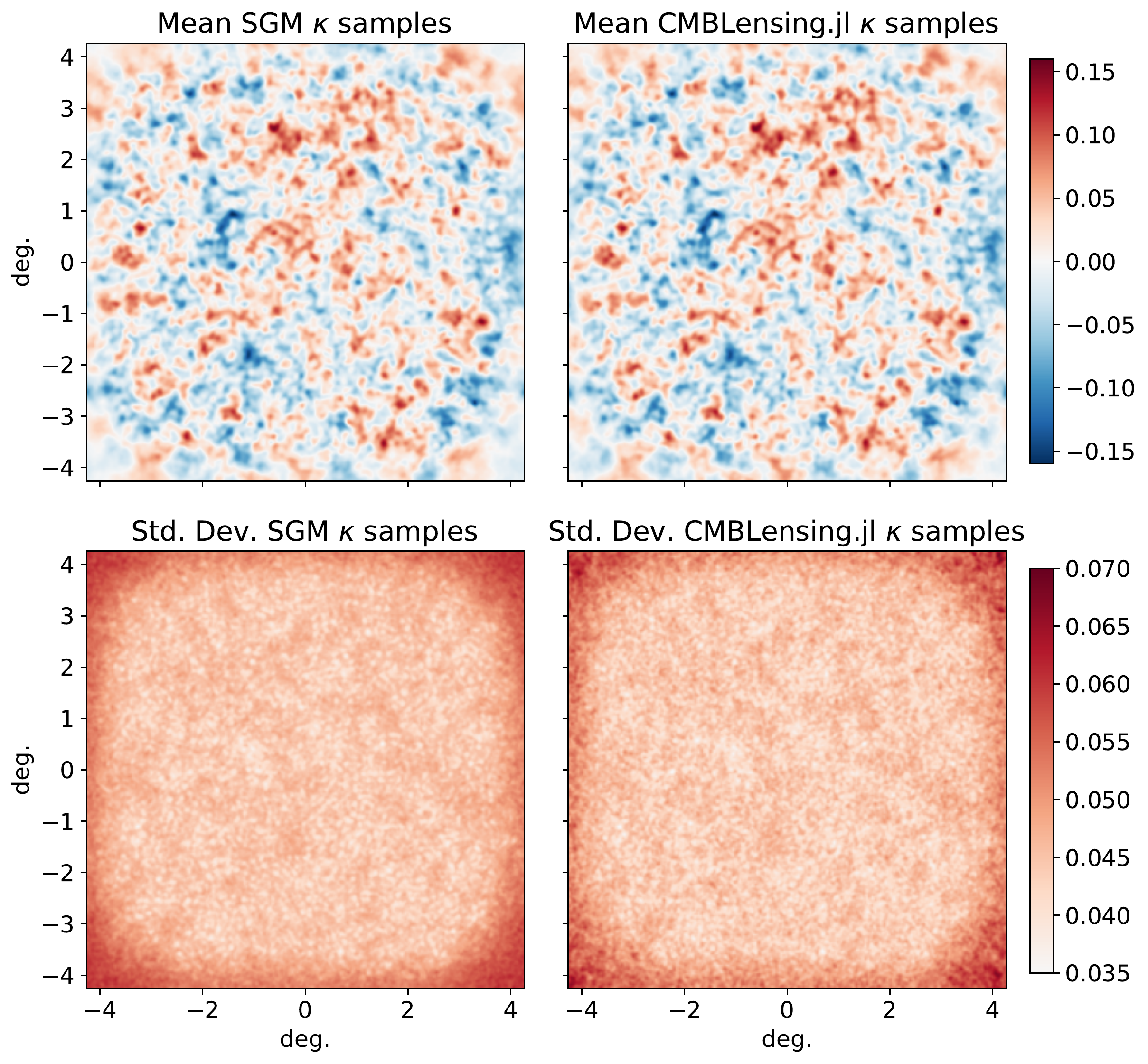}
    \caption{Results for the validation sample of the \cmbjl{} benchmark. \textbf{Top}: mean over 2048 posterior samples from the SGM (left) and the \cmbjl{} chain (right). \textbf{Bottom}: standard deviation of posterior samples from 1024 diffusion samples (left) and the \cmbjl{} chain (right).}
    \label{fig:1PARAMval}
\end{figure}

\begin{figure}
    \centering
    \includegraphics[scale=.6,left]{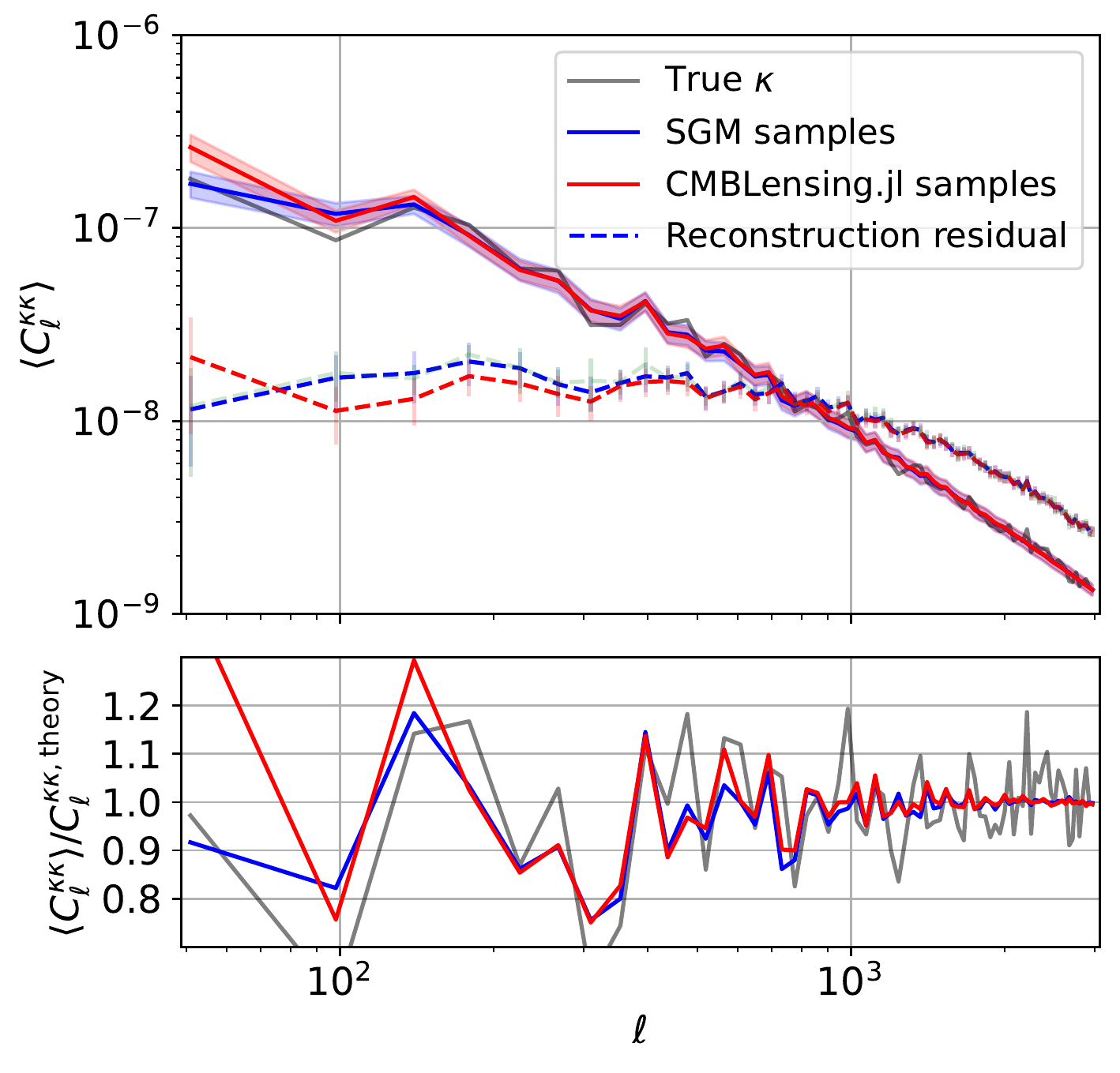}
    \caption{\textbf{Top:} Mean power spectrum of SGM samples (blue), \cmbjl{} samples  (red), and the true validation $\kappa$-map (gray). The reconstruction noise curves give the mean power spectrum of reconstruction residuals $(\kappa_{\rm sample} - \kappa_{\rm true})$.
    The shaded regions and error bars demark the $1\sigma$ spread of the power spectra. \textbf{Bottom:} mean power spectrum of samples divided by the theory power spectrum (at $A_\phi = 1$).}
    \label{fig:PS_1PARAM}
\end{figure}

\begin{figure}
    \centering
    \includegraphics[scale=.57,left]{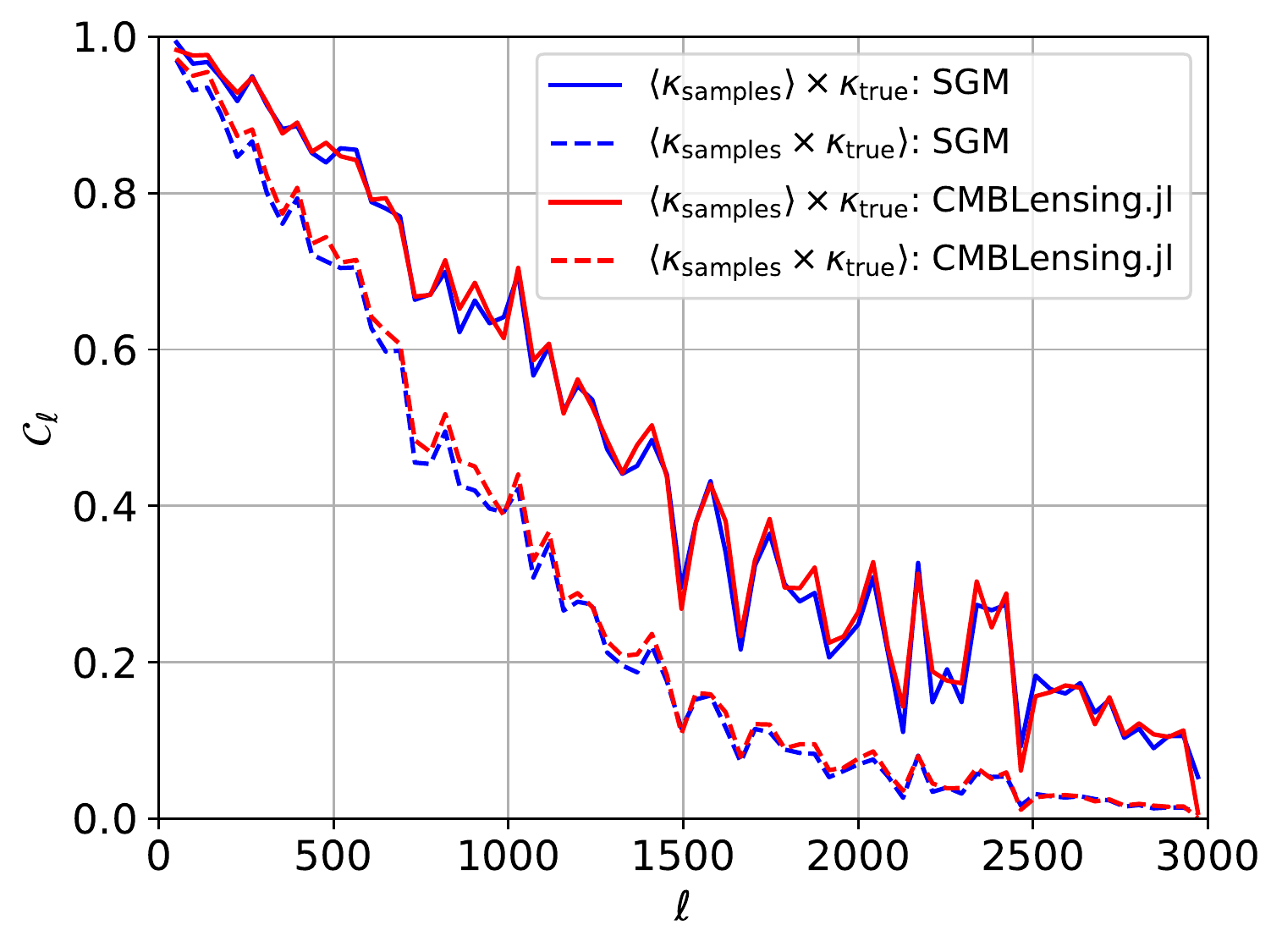}
    \caption{Cross-correlations (Eq. \eqref{eq:cross}) of validation samples with the true lensing convergence map. Solid lines denote cross-correlation of the mean of posterior samples from the SGM (blue) and \cmbjl{} chain (red). Dashed lines denote the mean of individual cross-correlations of samples.}
    \label{fig:C_1PARAM}
\end{figure}

We validate our generative model by comparing 2048 samples of our model and a \cmbjl{} chain of 17000 samples (we drop the first 1000 samples and keep every fifth sample, leaving us with 3200 samples from the chain), for a validation case with parameters $(A_\phi=1,\; r=0.04)$, shown in Fig. \ref{fig:validation_maps_1PARAM}. In Fig. \ref{fig:1PARAMval} we show the mean and standard deviation over posterior samples drawn using the two methods, while in Fig. \ref{fig:PS_1PARAM} we show the mean power spectrum of the samples. Additionally, we show the mean power spectrum of the reconstruction residual of samples $(\kappa_{\rm samples} - \kappa_{\rm true})$ in dashed lines, which quantifies the signal-to-noise ratio of the reconstruction on different scales. We see that for this configuration the reconstruction is signal-dominated up to $\ell \approx 800$. As a consequence, on small, noise-dominated scales the power spectrum of samples becomes increasingly prior driven, and the mean power spectrum of samples reduces to the theory spectrum, as observed in the bottom panel. We can further investigate the quality of our samples by comparing the correlation coefficient between samples and the true lensing convergence map as a function of scale:
\begin{eqnarray}
    \mathcal{C}_\ell = \frac{C^{\kappa_{\rm samples} \kappa_{\rm true}}_\ell}{\sqrt{C_{\ell}^{\kappa_{\rm true} \kappa_{\rm true}} C_{\ell}^{\kappa_{\rm samples} \kappa_{\rm samples}}}},
    \label{eq:cross}
\end{eqnarray}
shown in Figure \ref{fig:C_1PARAM}. The various metrics demonstrate good agreement between the \cmbjl{} chain samples and SGM samples. \\

We conclude that our score-based generative model has learned an accurate approximation of the lensing posterior, and can therefore be used to draw samples of the reconstructed lensing convergence. Our SGM model has the additional advantage that samples are uncorrelated, resulting in a more effective sampling of large angular scales compared to the \cmbjl{} chains, that struggle to converge here for masked data. In fact, for the \cmbjl{} chain presented here, the largest scale has not yet converged, explaining the large discrepancy of the largest scale bandpower with our SGM result, which more closely follows the truth.

\section{Non-Gaussian Lensing reconstruction}
\label{sec:NG}
Contrary to the MAW approach based on an explicitly specified posterior, our approach is based on a posterior that is implicitly specified through the training data. Our generative model therefore allows for a straightforward extension to non-Gaussian lensing potentials; we simply need to train our model on data consisting of non-Gaussian lensing maps and lensed CMB data. To demonstrate this, we train it using a set of simulated non-Gaussian lensing potentials provided by the authors of Ref.~\cite{Takahashi:2017hjr}. This dataset consists of $108$ high-resolution full-sky lensing convergence maps generated by ray tracing dark matter N-body simulations.\\

We use $100$ of their full-sky $\kappa$ maps to lens $100$ independent full-sky $(Q,U)$-maps using the lensing algorithm implemented in the \texttt{lenspyx} code \cite{Reinecke:2023gtp}. We generate $256$ flat-sky observations (and the corresponding patches of the lensing convergence) per full-sky map, with the same properties as before (i.e. Table \ref{tab:config}). These $25600$ flat-sky patches serve as training data for our model, while one of the remaining 8 full-sky maps is used to generate independent validation data. We keep cosmological parameters fixed ($A_\phi=1,\; r=0$) for all data for simplicity, but in principle, $r$ can be varied in the training data as in the previous section.\\

\begin{figure}
    \centering
    \includegraphics[scale=.355]{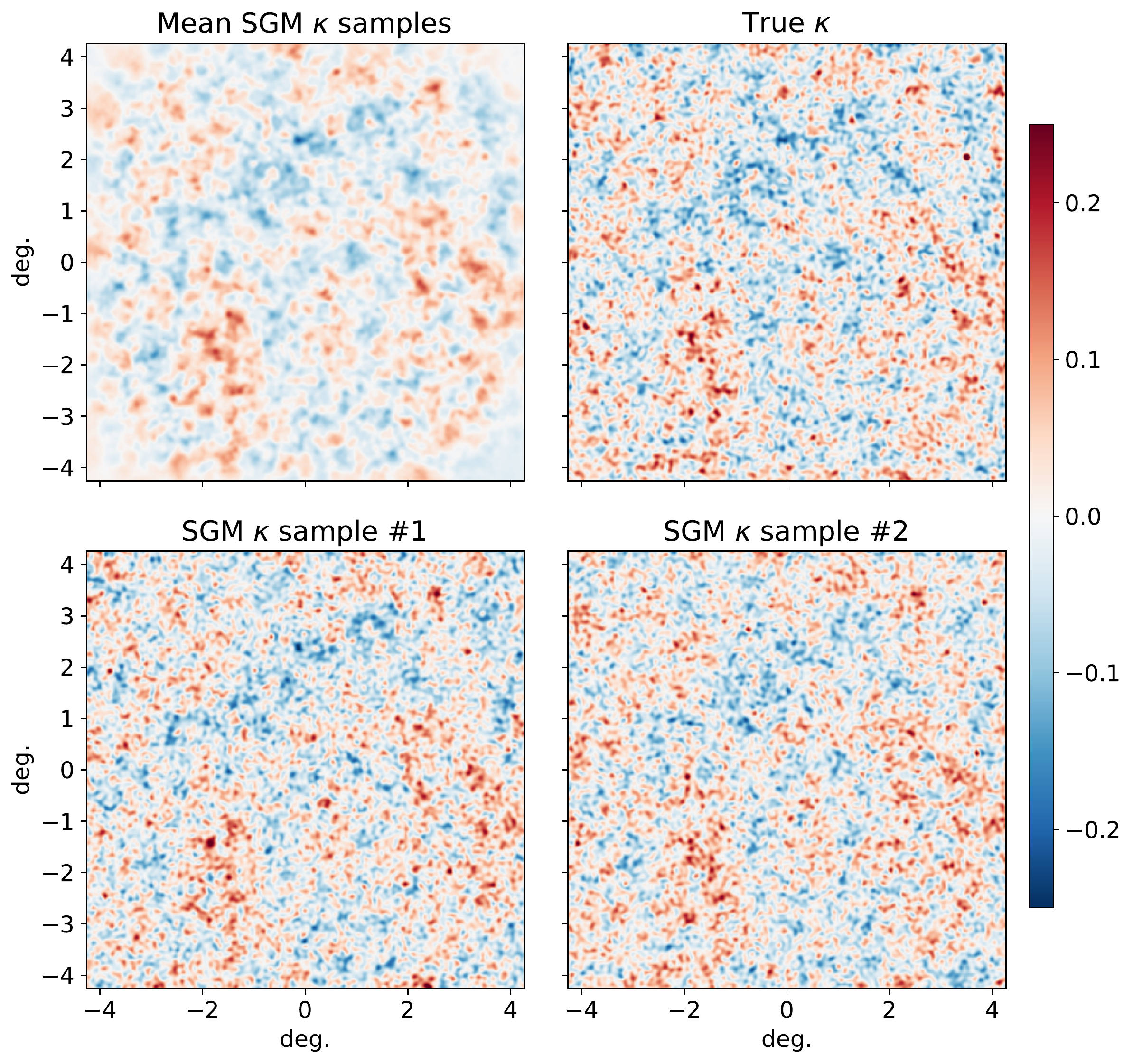}
    \caption{Results for the validation case in the N-body non-Gaussian setup. \textbf{Top left:} mean over 4096 posterior samples. \textbf{Top right:} the true lensing convergence. \textbf{Bottom:} two independent samples of the convergence map. }
    \label{fig:NGval}
\end{figure}

\begin{figure}
    \includegraphics[scale=.6,left]{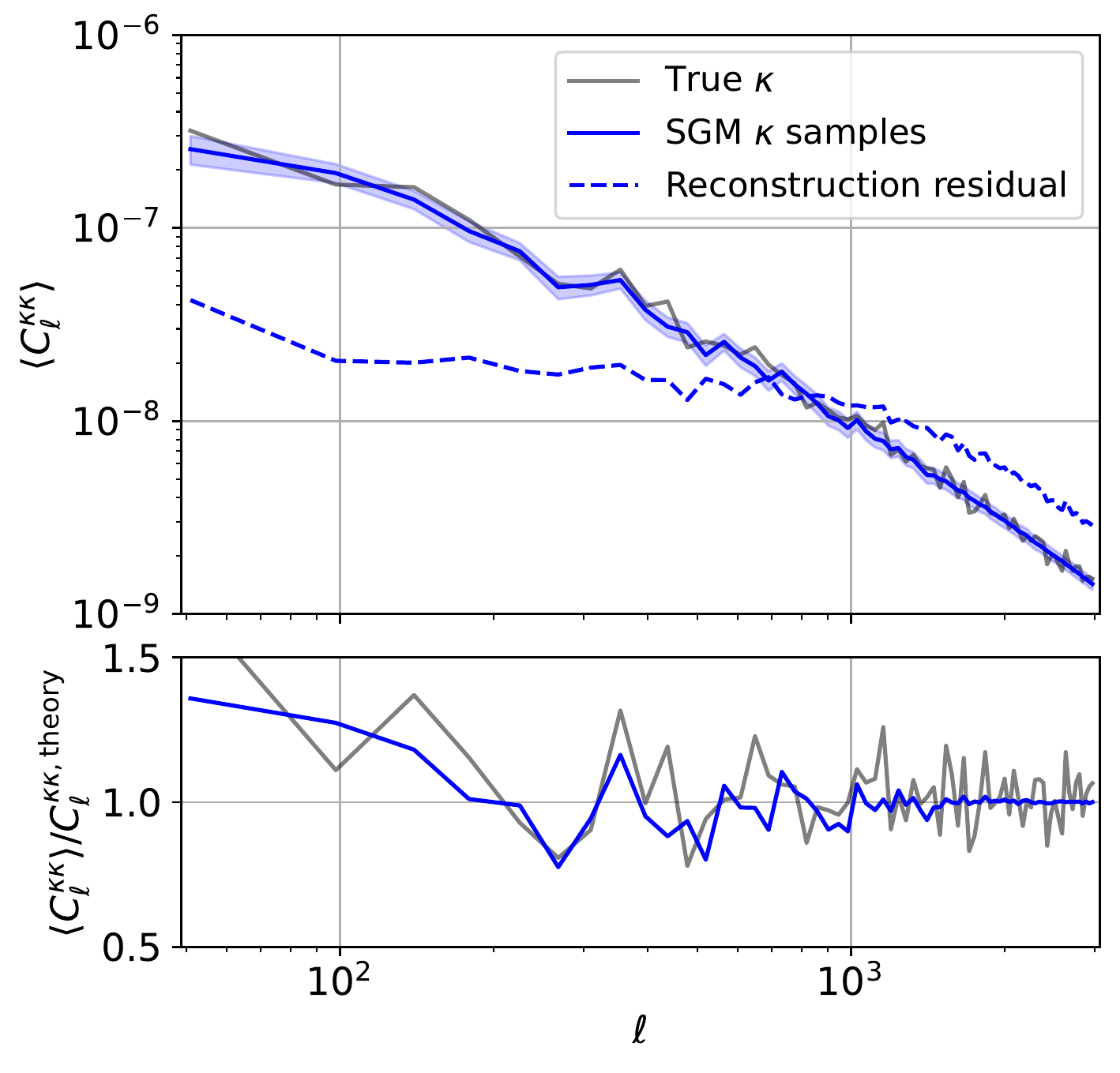}
    \caption{\textbf{Top:} Mean power spectrum of SGM samples for the N-body validation case presented in Figure \ref{fig:NGval}. The reconstruction noise curve gives the mean power spectrum of reconstruction residuals $\kappa_{\rm sample} - \kappa_{\rm true}$. \textbf{Bottom:} mean power spectrum of samples divided by the theory power spectrum (at $A_\phi = 1$).}
    \label{fig:NGPS}
\end{figure}

\begin{figure}
    \includegraphics[scale=.57,left]{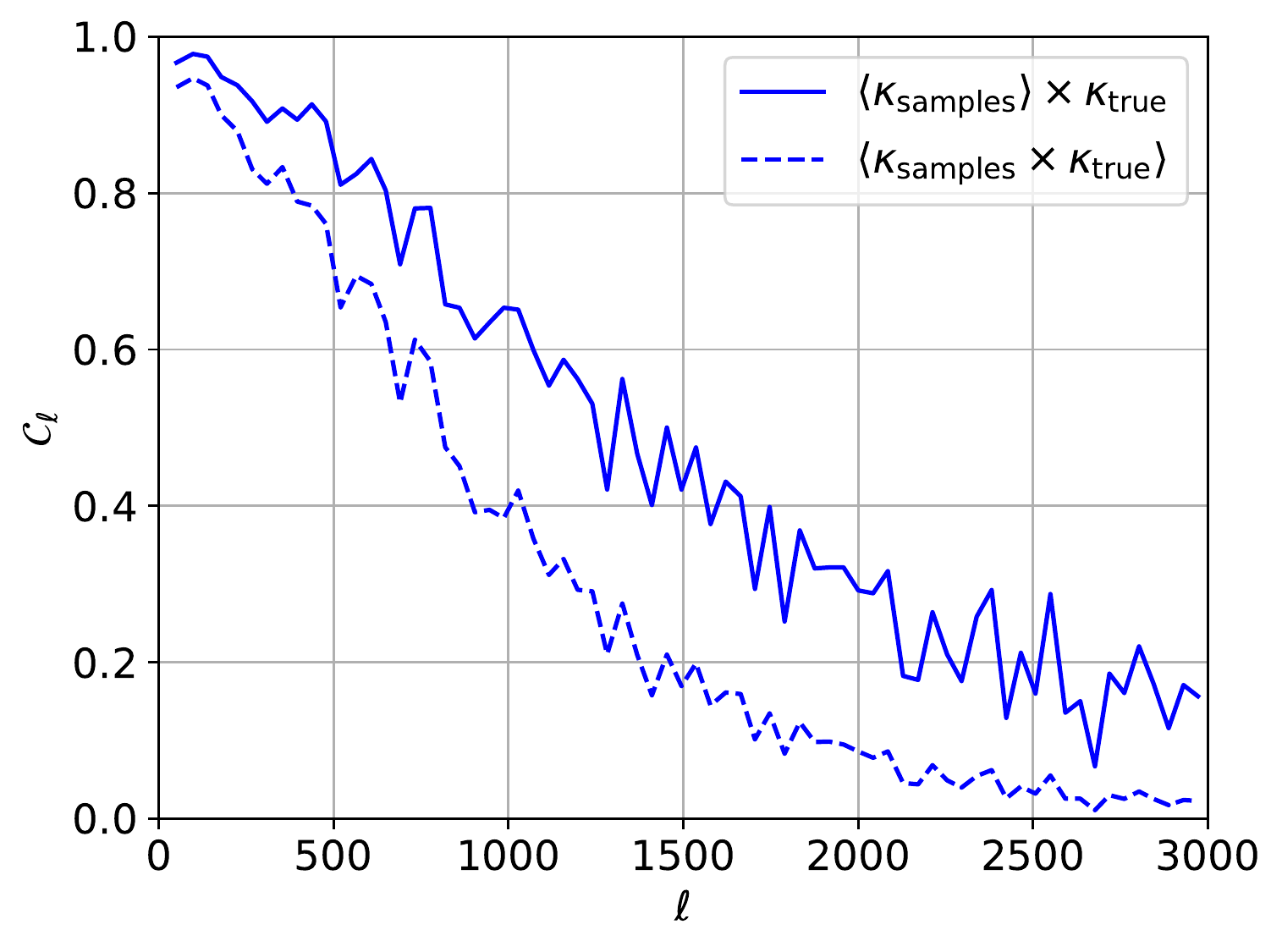}
    \caption{Cross-correlations of validation samples with the true lensing convergence map for the non-Gaussian case. Solid lines denote the cross-correlation of the mean of posterior samples from the SGM. Dashed lines denote the mean of individual cross-correlations of samples.}
    \label{fig:NGcross}
\end{figure}

Once trained, we draw 4096 samples from our model for a validation case, the results of which are shown in Figure \ref{fig:NGval}. Both the samples and the mean over posterior samples show a clear correlation with the true non-Gaussian $\kappa$ map. The power spectrum and cross-correlation coefficient for the non-Gaussian validation case are shown in Figure \ref{fig:NGPS} and Figure \ref{fig:NGcross}.\\

Different from the Gaussian case of the previous section, the non-Gaussian lensing convergence maps contain information beyond the power spectrum. Such non-Gaussian statistics of the lensing field have been studied previously in \cite{Liu:2016nfs}, in particular the one-point probability distribution function and peak counts. Additionally, we can study higher-order correlation functions, such as the bispectrum \cite{Namikawa:2016jff}.\\

\begin{figure}
    \vspace{.1cm}
    \includegraphics[scale=.57,left]{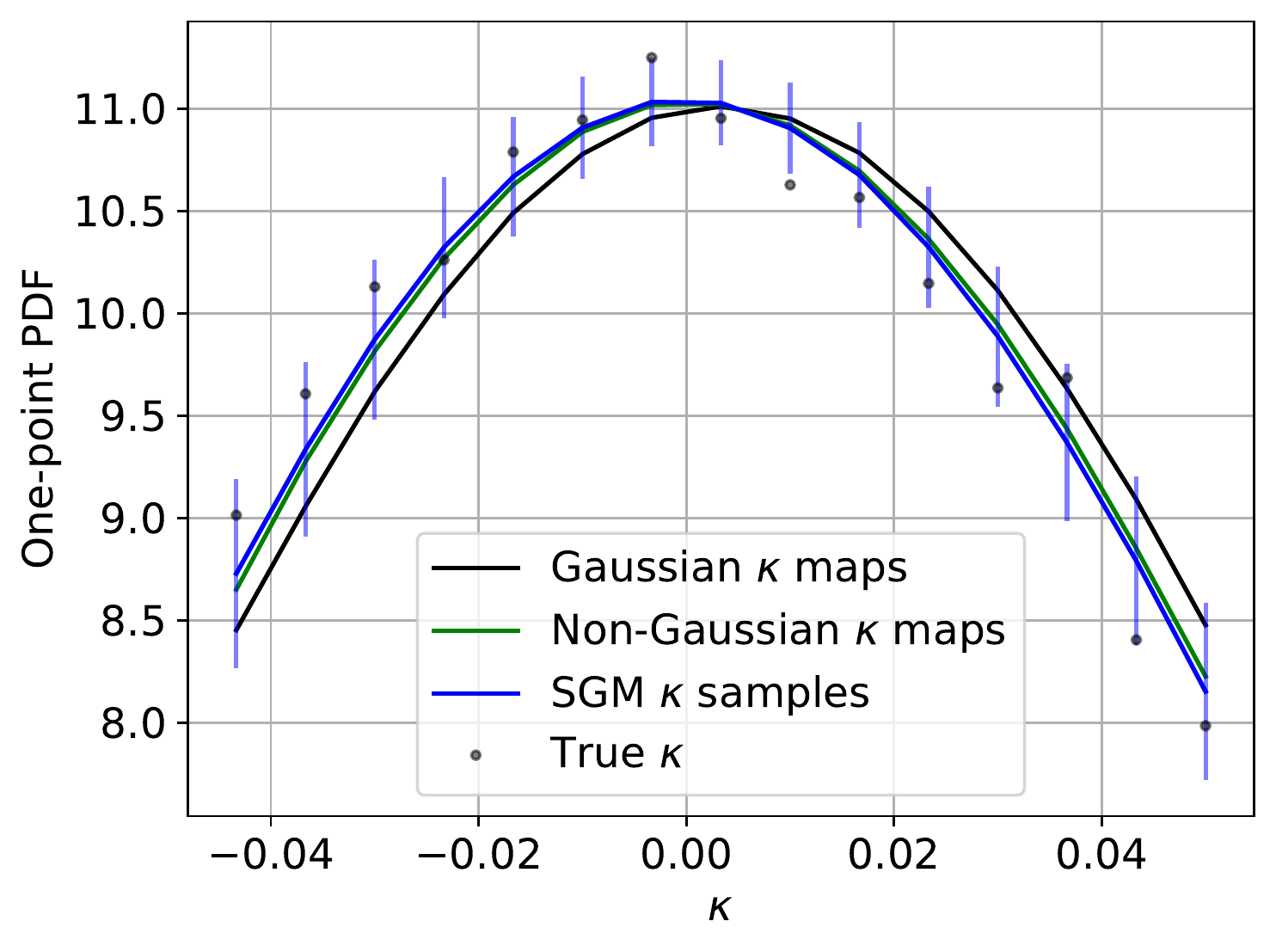}
    \caption{One-point probability distribution function of $\kappa$, using 15 linear bins between $\kappa \in (-0.05,0.05)$. In green, we show the PDF of a set of the N-body non-Gaussian $\kappa$ maps. In black, we show the PDF of a set of Gaussian $\kappa$ maps generated with the same power spectrum as the non-Gaussian maps. In blue, we show the PDF of a set of samples using our generative model, and the error bars denote its standard deviation per bin. Gray dots denote the one-point PDF of the true validation $\kappa$ map.}
    \label{fig:NG1PDF}
\end{figure}

In Figure \ref{fig:NG1PDF} we show the one-point PDF of validation $\kappa$ samples, and that of both Gaussian and Non-Gaussian $\kappa$ maps, clearly showing that our model produces one-point statistics that are consistent with the mean over many non-Gaussian maps, but slightly shifted towards the statistics of the true $\kappa$ map.\\

\begin{figure*}[htbp]
    \centering
    \begin{minipage}{0.49\textwidth}
        \centering
        \includegraphics[width=\linewidth]{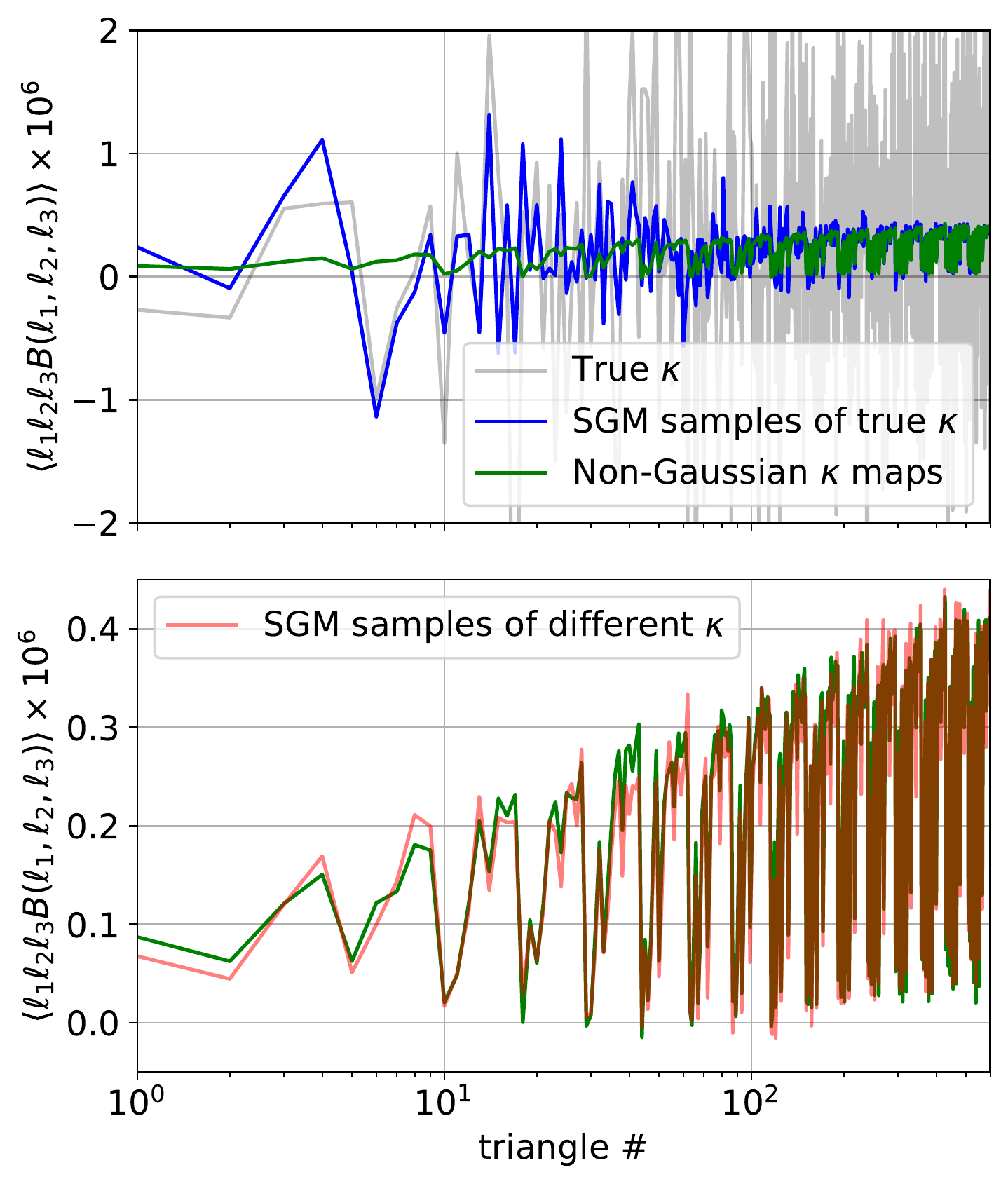}
    \end{minipage}\hfill
    \begin{minipage}{0.49\textwidth}
        \centering
        \includegraphics[width=\linewidth]{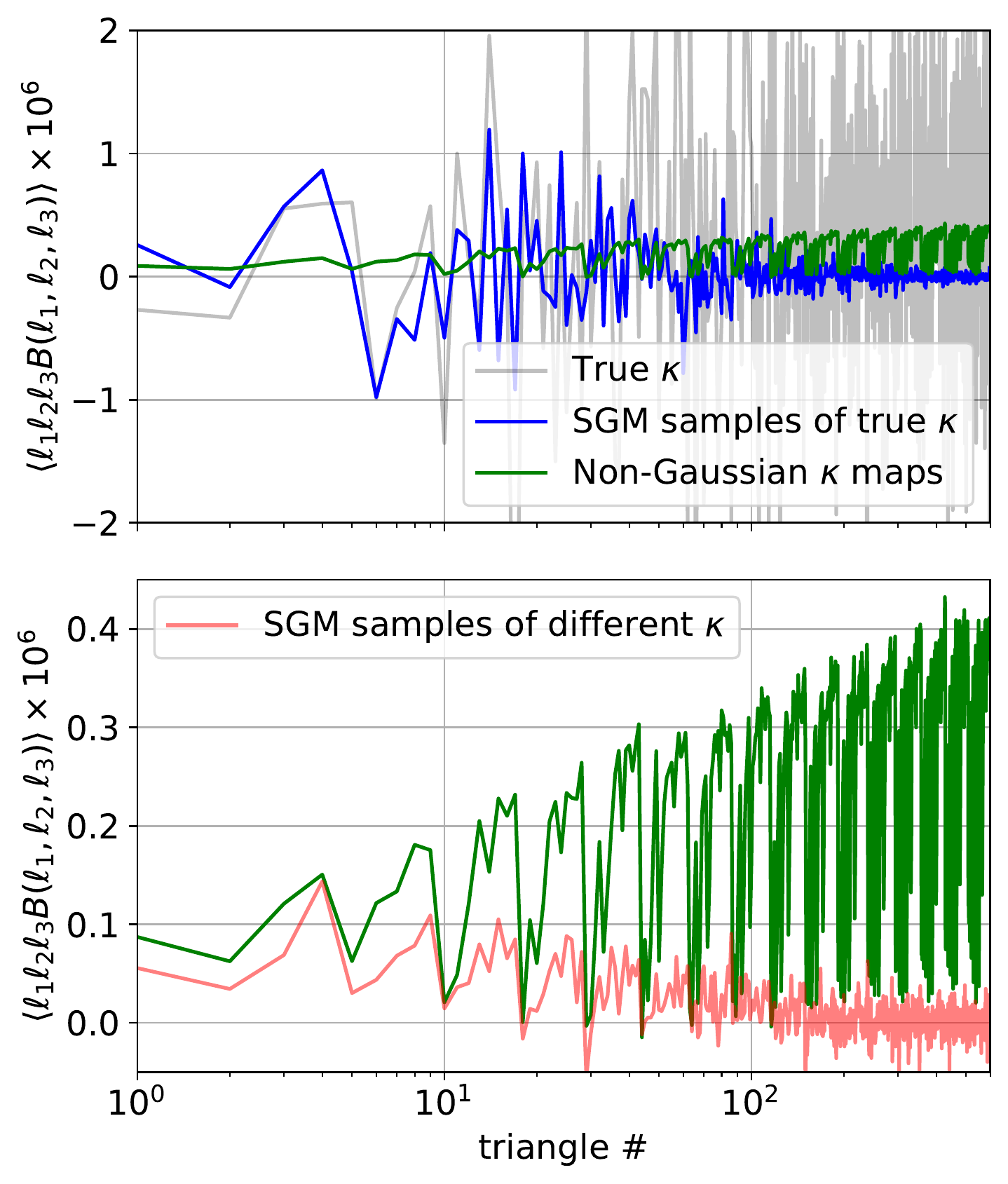}
    \end{minipage}
    \caption{Bispectrum results for an SGM trained on non-Gaussian (left) or Gaussian (right) lensing data. \textbf{Top:} bispectrum of the true validation lensing convergence $\kappa$ map (gray), and mean bispectrum of many different non-Gaussian $\kappa$ maps (green), and SGM samples of the true $\kappa$ map (blue). \textbf{Bottom:} mean bispectrum of SGM samples of many different $\kappa$ maps (red). The green line is an approximation to the theory bispectrum.}
    \label{fig:B_full_comparison}
\end{figure*}

Next, we investigate the bispectrum of the samples. We compute the bispectrum using 17 bins with a width of four times the largest mode, $\ell_{\rm min} \approx 43$, of the flat-sky patches, resulting in 597 triangle configurations up to $\ell_{\rm max} \approx 2911$. Figure \ref{fig:B_full_comparison}a shows the mean bispectrum of samples and many N-body non-Gaussian maps. We plot the triangle configurations logarithmically to emphasize triangles on large scales, where the reconstruction is signal-dominated (as in Figure \ref{fig:NGPS}). On large scales (low triangle index), the mean bispectrum of samples (blue) follows the bispectrum of the true lensing convergence map (gray) more closely than it follows the theory bispectrum of non-Gaussian maps (green). On smaller scales, where the reconstruction is noise-dominated, the bispectrum of samples more closely follows the theory bispectrum (i.e. falls back to the learned non-Gaussian prior).  These results show that our model has indeed learned a non-Gaussian posterior and prior and generates accurate samples of the non-Gaussian lensing convergence map.\\

To demonstrate the error induced by drawing samples from the less informed posterior (with Gaussian lensing prior) we show the bispectrum for samples of the validation case drawn from a model that was trained using only Gaussian lensing maps (with the same power spectrum as the non-Gaussian N-body lensing maps). In Figure \ref{fig:B_full_comparison}b we show the mean bispectrum of 4096 validation samples from this Gaussian model. We expect such a model to be able to capture non-Gaussian information only on highly signal-dominated regions where the posterior is not dominated by the Gaussian prior (that is implicitly learned from the Gaussian training data). Indeed the top panel shows that on the largest, signal-dominated scales, the model produces a bispectrum that follows the true validation map and is similar to that of the SGM trained on non-Gaussian maps. However, on smaller, noise-dominated scales the bispectrum vanishes, as the model falls back on the internally learned Gaussian prior, unlike our non-Gaussian model. Furthermore, in the bottom panel, we see that the mean bispectrum over 4096 samples of different lensing realizations, does not closely follow the theory bispectrum.


\section{Conclusion \& Outlook}
In this work, we demonstrated the use of score-based generative models in CMB lensing reconstruction tasks. We have shown that such models can effectively learn the conditional posterior distribution of the lensing convergence map given observed CMB data, by training on corresponding pairs of these. The trained model can be used to rapidly draw lensing posterior samples. For the case of Gaussian lensing convergence maps, we have validated our methodology by comparing the results of our model against the established Hamiltonian Monte Carlo methods of Millea, Anderes and Wandelt \cite{Millea:2020iuw}.\\

Since our model learns the posterior distribution entirely from simulated data, and the prior and likelihood are modeled through the processes encoded in the simulations, it can be used in situations where the analytic form of the posterior is not known, such as in the case of non-Gaussian lensing convergence maps. Using realistic non-Gaussian lensing maps obtained by ray-tracing N-body simulations, we demonstrate that our model can indeed be trained to yield samples with the correct non-Gaussian statistics (i.e. one-point probability distribution and bispectrum). As we have shown, the resulting samples of our model can be used to estimate the CMB lensing bispectrum, which is a difficult task using traditional estimators such as the quadratic estimator, because of large noise biases \cite{Kalaja:2022xhi}.\\

Another important advantage of our model as compared to the HMC method is speed and efficiency. Our model, with neural network configuration as described in Appendix \ref{App:UNET} and using 1000 denoising steps, generates a single sample in 12 seconds, which can be vectorized up to 128 simultaneous samples in about 225 seconds ($\sim 1.75$ seconds per sample) on an A100 GPU, and can be easily parallelized over multiple GPUs. Significant speed improvements can be expected from optimizing neural network complexity and noise schedule (i.e. number of denoising steps), which is beyond the scope of this paper. Notably, the samples of our model are completely uncorrelated. Meanwhile, drawing a single sample with \cmbjl{} (in the configuration used in this paper, varying only r) takes $\sim 1.6$ seconds on an A100 GPU. However, HMC samples are correlated. Taking the autocorrelation length of the bandpower on the least efficiently sampled scale of the chain (which is found to be about $290$), shows that the effective independent sample size of the chain is more than two orders of magnitude smaller. It should be noted that improving the HMC analysis with suitable approximations used in real lensing analyses, is likely to again yield an order of magnitude improvement in speed for the configuration considered in this work \cite{Millea:2021had,millea2022improved}. Finally, for larger maps, the HMC analysis requires more samples to converge, thus scaling inefficiently with problem size. For our model, besides the obvious increase in training and sampling time due to the larger data size, there is no such chain convergence issue, thus promising more efficient scaling to larger problems. We conclude that generative models allow for a more efficient approach to Bayesian CMB lensing reconstruction, which translates to faster convergence of analysis pipelines. \\

Looking ahead, ideally, the model would include the amplitude of the CMB lensing power spectrum as well. As mentioned earlier, we have been unsuccessful at this so far. Nevertheless, we are optimistic that with an appropriate modification of the model or algorithm, this can be overcome. We share some of our thoughts on this issue in Appendix \ref{app:amp}. Alternatively, sampling the cosmological parameters can also be left as a seperate Gibbs-like steps using another simulation-based inference pipeline conditioned on the data and our model's lensing samples. We leave an investigation into this for future work. \\

Though in this work we have focused on the non-Gaussianity of the lensing potential, our model's flexibility straightforwardly allows for other extensions that would similarly complicate the posterior evaluation, such as the inclusion of inhomogeneous and anisotropic noise in the data \cite{Hanson_2009}. Once again, this would require only a modification of the simulated data that the model is trained on. Along these lines, obvious directions for future work are to assess biases due to foregrounds \cite{vanEngelen:2012va,Osborne:2013nna,ACT:2023ubw} and correlated point-source masks \cite{Lembo:2021kxc} that complicate the reconstruction of the lensing potential, and reconstruction of lensing curl modes \cite{Cooray:2005hm,Namikawa:2011cs,Namikawa:2013wda,Pratten:2016dsm,Robertson:2023xkg}.\\

The results of this work pave the way for more efficient, accurate, and realistic lensing reconstruction. The resulting samples of our model can be used in subsequent analyses of the CMB, or as an unbiased tracer of the total matter density distribution while being able to appropriately propagate errors due to uncertainty in the lensing reconstruction.\\

\noindent Our model and the code used to generate the results of this paper is publicly available at \url{https://github.com/tsfloss/CMBLensingDiffusion}

\section{Acknowledgements}
The authors would like to thank Julien Carron, Shirley Ho, François Lanusse, Ronan Legin, Florian List, Daan Meerburg, Marius Millea, and Sebastian Wagner-Carena for useful discussions. We furthermore thank Marius Millea for providing support with \cmbjl{}. T.F. thanks the Flatiron Institute's Center for Computational Astrophysics, where part of this research was conducted, for its hospitality. We thank the Center for Information Technology of the University of Groningen for providing access to the Hábrók high-performance computing cluster. T.F. is supported by the Fundamentals of the Universe research program at the University of Groningen. B.D.W. acknowledges support from the DIM ORIGINES project "INFINITY NEXT - Calcul Hybride et IA en Astrophysique". The Flatiron Institute is supported by the Simons Foundation. 

\appendix
\section{Neural Network Specifics}
\label{App:UNET}
The U-Net used in our score-based generative model employs double convolution blocks, each of which consists of a $3\times3$ convolution with 64 channels, a sinusoidal time embedding, group normalization, and SiLU activation, followed (preceded) by a Fourier pooling (upsampling) layer in the encoding (decoding) process. The Fourier pooling layer performs the downsampling of the image in the Fourier domain, conserving the data's frequency content more accurately than typical pooling or stride convolutions. Similarly, the Fourier upsampling (or interpolation) layer upsamples the data in the Fourier domain, by filling up the modes of a higher-resolution image with the modes of the lower-resolution image, again exactly preserving the frequency content. We furthermore employ residual connections (rescaled with $\sqrt{2}$) and skip-connections. At the network bottleneck, the data has a size of $8^2$ pixels. The network input is a channel-wise concatenation of the noised image $\mathbf{x}_i$ and conditioning data $\mathbf{y}$ \cite{batzolis2021conditional}. The model is trained using the AdamW optimizer with gradient norms clipped to $1$, and a batch size of $128$. We first train with a learning rate of $0.0002$, followed by a learning rate of $10^{-5}$ to ensure convergence. Additionally, we use exponential moving average (EMA) weights, with a decay rate of $0.9999$, to further stabilize the model during inference. Each model is trained for a total of around $200000$ steps, taking around $16$ hours on two A100 GPUs. Our model is implemented using JAX \cite{jax2018github}.

\section{Varying the lensing amplitude}
\label{app:amp}
\begin{figure}
    \includegraphics[scale=.59,left]{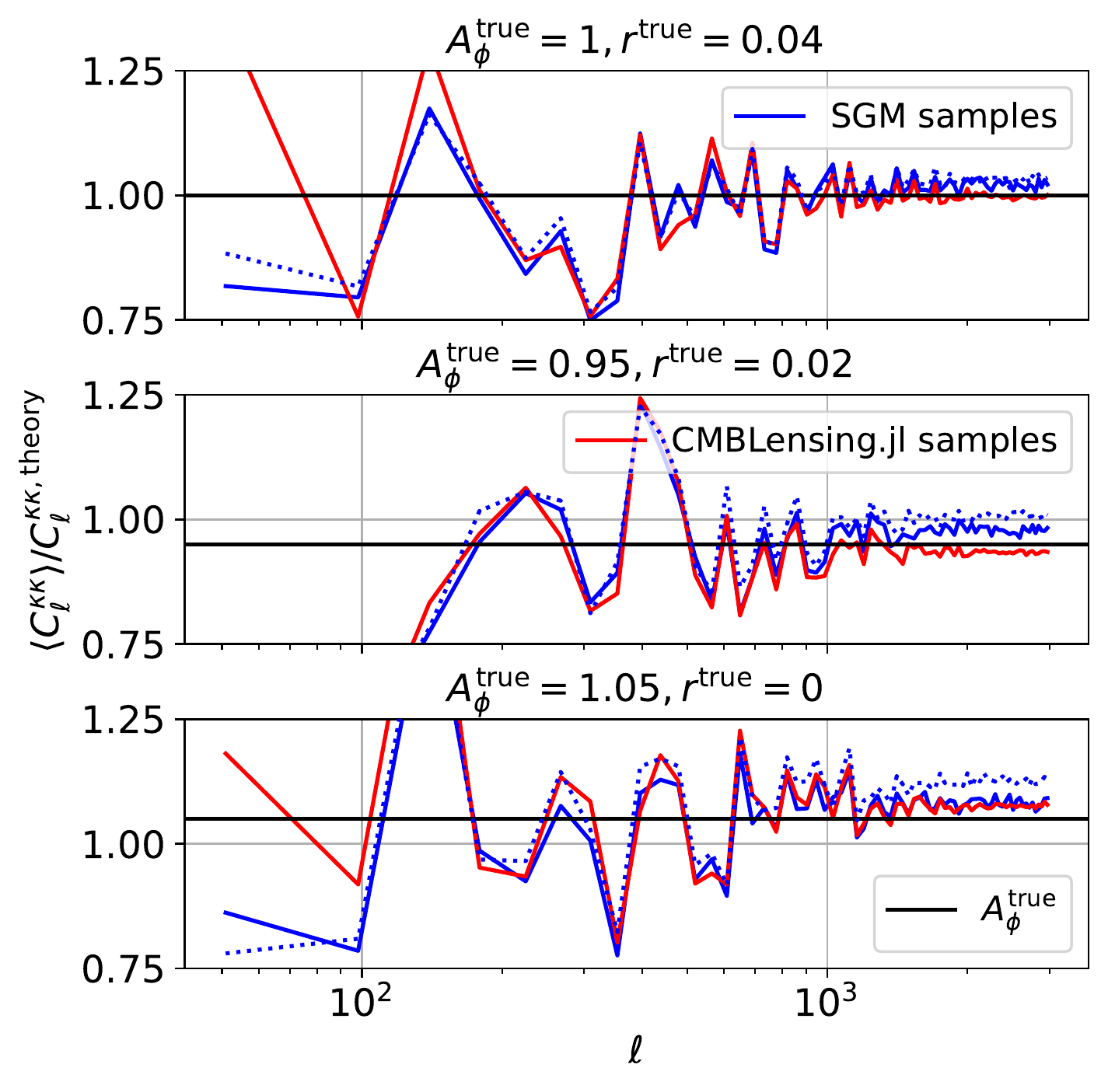}
    \caption{Mean power spectrum, normalized by the theory power spectrum at $A_\phi = 1$, of samples from different models (solid and dotted blue) and \cmbjl{} chains (red). for three different realizations and parameters.}
    \label{fig:PS_2PARAM}
\end{figure}

In Section \ref{sec:benchmark}, we only varied the tensor-to-scalar ratio $r$. Ideally, we would like to be able to marginalize over the amplitude $A_\phi$ of the lensing power spectrum too. However, we found that the diffusion algorithm as is, is unable to yield samples with the correct amplitude on small scales in this case, even though the phases of the samples are accurate (as quantified by the cross-correlation coefficient in equation \eqref{eq:cross}, that does not contain the amplitude). We find that distinct models converge to different results. An example of this is visualized in Figure \ref{fig:PS_2PARAM}, where we show results of the mean power spectrum of samples for three different realizations and parameters, from two models (solid and dotted) trained on data that simultaneously vary $A_\phi$ and $r$, compared to those of \cmbjl{} chains. These two models have also been trained for a significantly longer time than the models presented in the main text, to ensure convergence of their EMA weights. \\

Although we are uncertain as to what exactly causes this issue, we think that it could be related to the nature of the diffusion sampling algorithm. The algorithm assumes that the signal-to-noise ratio of the noised image at different timesteps follows the noise schedule $\sigma(t)$ in equation \eqref{eq:schedule}. However, when varying the amplitude of the underlying data, at a fixed time this ratio varies significantly among the data, especially at lower noise levels. Meanwhile, during the sampling process, at every denoising step, the network makes a guess for the amplitude that will not be perfect. Thus, at the next timestep, there is a mismatch between the actual signal-to-noise ratio and that expected by the model. The accumulation of many of these mismatches could the wrong amplitude, especially on noise-dominated scales (where the value of the amplitude is most relevant), even though the phase structure is accurately resolved. From this perspective, a blind denoising approach, such as that presented in \cite{Heurtel-Depeiges:2024ebo}, might prove useful.

\bibliographystyle{apsrev4-1}
\bibliography{bibliography}

\end{document}